\documentclass[preprint,prd,amsmath,amssymb,aps,nofootinbib]{revtex4-1}

\usepackage{graphicx}
\usepackage{bm}
\usepackage{natbib}
\def\mathbi#1{\textbf{\em #1}}

\def\aap{Astron.\ Astrophys.\ }
\def\apj{Astrophys.\ J.\ }

\def\mnras{Mon.\ Not.\ R.\ Astron.\ Soc.\ }

\def\ssr{Space\ Sci.\ Rev.\ }

\begin{document}
\title{Self-consistent interpretations of the multi-wavelength gamma-ray 
spectrum of LHAASO J0621$+$3755}
\author{Kun Fang$^a$} \email{fangkun@ihep.ac.cn}
\author{Shao-Qiang Xi$^a$} \email{xisq@ihep.ac.cn}
\author{Xiao-Jun Bi$^{a,b}$}

\affiliation{
 $^a$Key Laboratory of Particle Astrophysics, Institute of High Energy
 Physics, Chinese Academy of Science, Beijing 100049, China \\
 $^b$University of Chinese Academy of Sciences, Beijing 100049, China\\
 }

\date{\today}

\begin{abstract}
LHAASO~J0621$+$3755 is a TeV gamma-ray halo newly identified by LHAASO-KM2A. It is likely to be generated by electrons trapped in a slow-diffusion zone around PSR~J0622$+$3749 through inverse Compton scattering. When the gamma-ray spectrum of LHAASO-KM2A is fitted, the GeV fluxes derived by the commonly used one-zone normal diffusion model for electron propagation are significantly higher than the upper limits (ULs) of Fermi-LAT. In this work, we respectively adopt the one-zone superdiffusion and two-zone normal diffusion models to solve this conflict. For the superdiffusion scenario, we find that a model with superdiffusion index $\alpha\lesssim1.2$ can meet the constraints of Fermi-LAT observation. For the two-zone diffusion scenario, the size of the slow-diffusion zone is required to be smaller than $\sim50$ pc, which is consistent with theoretical expectations. Future precise measurements of the Geminga halo may further distinguish between these two scenarios for the electron propagation in pulsar halos.
\end{abstract}

\maketitle

\section{Introduction}
\label{sec:intro}
Pulsar halos, i.e., extended TeV gamma-ray emission around middle-aged pulsars, are believed to be a new class of gamma-ray sources \cite{Linden:2017vvb,Sudoh:2019lav,Giacinti:2019nbu}. These halos are generated by free electrons and positrons\footnote{\textit{Electrons} will denote both electrons and positrons hereafter.} escaping from the corresponding pulsar wind nebulae (PWNe) and wandering in the interstellar medium (ISM). The surface brightness profile (SBP) of the Geminga halo measured by HAWC constrains the diffusion of particles away from the pulsar to be much slower than that in the typical ISM \cite{Abeysekara:2017old}. This anomalously slow diffusion arouses extensive discussions on how particles propagate in pulsar halos \cite{Evoli:2018aza,Kun:2019sks,Liu:2019zyj,Wang:2021xph,Recchia:2021kty} and whether nearby pulsars can contribute significant positron flux at Earth \cite{Hooper:2017gtd,Fang:2018qco,Profumo:2018fmz,Tang:2018wyr,Shao-Qiang:2018zla,DiMauro:2019yvh,Fang:2019ayz}.

Recently, the LHAASO collaboration reports an extended TeV gamma-ray source named LHAASO J0621$+$3755, which is very likely to be a new pulsar halo \cite{Aharonian:2021jtz}. The associated pulsar, PSR~J0622$+$3749, is located right in the center of the gamma-ray halo and has a similar age and spin-down luminosity to Geminga. Meanwhile, the GeV observation of Fermi-LAT does not find extended emission around the pulsar and flux upper limits (ULs) can be obtained. However, assuming the commonly used one-zone normal diffusion (normal diffusion for short) model for electron propagation, the GeV fluxes extrapolated from the LHAASO-KM2A observation are significantly higher than the ULs of Fermi-LAT, unless an extreme injection spectrum is assumed (see Fig.~S4 of the Supplemental Material of Ref.~\cite{Aharonian:2021jtz}).

The normal diffusion model is not the only possible scenario to describe the electron transport in the pulsar halos. Multi-scale inhomogeneities may exist in the ISM, and the normal diffusion could be generalized to superdiffusion. The superdiffusion model has been applied in different fields of astrophysics to solve specific problems \cite{1998AdSpR..22...55V,2001NuPhS..97..267L,2015JPhCS.632a2027V,2016A&A...596A..34P, 2018MNRAS.478.4922Z}. We have tested the superdiffusion model by fitting the SBP of the Geminga halo and found that it is permitted by the observation of HAWC \cite{Wang:2021xph}. An important character of superdiffusion is that it can predict much higher electron flux at large distance from the source than that of the normal diffusion. We have found that Geminga can contribute considerable positron flux at Earth under the superdiffusion model even if the small diffusion coefficient around Geminga is extrapolated to the whole region between Geminga and the Earth. We will show below that the conflict between the TeV and GeV observations for LHAASO~J0621$+$3755 could be solved in the superdiffusion scenario.

Another possible solution to this problem is the two-zone diffusion model \cite{Hooper:2017gtd,Fang:2018qco}. The significant inconsistency between the diffusion coefficients in the pulsar halos and the average coefficient of the Galaxy indicates that the slow diffusion around the pulsars should not be typical in the Galaxy. Considering the possible origins \cite{Evoli:2018aza,Kun:2019sks}, the slow diffusion may only exist in the nearby region of the pulsars ($\lesssim100$ pc). As shown in Ref.~\cite{Aharonian:2021jtz}, the two-zone model can explain the spectrum in the energy range from a few tens of GeV to $\sim$ 100 TeV.

In this work, we attempt to consistently explain the TeV and GeV gamma-ray observations of LHAASO~J0621$+$3755 with the two models described above, respectively. In Sec.~\ref{sec:method}, we introduce the electron propagation, which is the core of the calculation of the gamma-ray SBP and energy spectrum. As the Fermi-LAT ULs are model-dependent, we introduce the analysis of the Fermi-LAT data in Sec.~\ref{sec:fermi}. In Sec.~\ref{sec:super}, we fit the SBP measured by LHAASO-KM2A and explain the multi-wavelength gamma-ray spectrum with the one-zone superdiffusion (superdiffusion for short) model. In Sec.~\ref{sec:2zone}, we adopt the two-zone normal diffusion (two-zone diffusion for short) model to explain the observations and constrain the size of the slow-diffusion zone. The conclusion is in Sec.~\ref{sec:conclude}.

\section{Electron propagation}
\label{sec:method}
To get the gamma-ray SBP and energy spectrum of the pulsar halo, we solve the electron propagation equation to obtain the electron number density around the pulsar and then do the line-of-sight integration to get the electron surface density. The electrons emit the gamma rays through the inverse Compton scattering (ICS). We adopt the standard formula given in Ref.~\cite{Blumenthal:1970gc} to calculate the ICS. In the following, we introduce the calculation of electron propagation for both the superdiffusion and two-zone diffusion models.

\subsection{Propagation equation}
\label{subsec:trans}
Electrons are continuously scattered by the chaotic magnetic field in the ISM 
after being injected from the PWN. The general electron propagation equation 
for both the superdiffusion and two-zone diffusion scenarios can be expressed by
\begin{equation}
  \frac{\partial N(E_e, \mathbi{r}, t)}{\partial t} = -D(E_e, \mathbi{r}, 
\alpha)(-\Delta)^{\frac{\alpha}{2}} N(E_e, \mathbi{r}, t) +
\frac{\partial[b(E_e)N(E_e, \mathbi{r}, t)]}{\partial E_e} + Q(E_e,
\mathbi{r}, t)\,,
 \label{eq:prop}
\end{equation}
where $N$ is the electron number density and $E_e$ is the electron energy. The 
superdiffusion exponent is denoted by $\alpha$, the domain of which is $(0, 
2]$. When $\alpha=2$, the propagation degenerates to the normal diffusion. The 
diffusion coefficient $D$ is assumed to have an energy dependency of 
$D\sim E_e^{1/3}$, which is predicted by the Kolmogorov's theory. For the 
two-zone diffusion case, the diffusion coefficient is written as
\begin{equation}
 D(E_e, \mathbi{r}, 2)=\left\{
 \begin{aligned}
  & D_1(E_e),\quad |\mathbi{r}-\mathbi{r}_s|<r_\star \\
  & D_2(E_e),\quad |\mathbi{r}-\mathbi{r}_s|\geq r_\star \\
 \end{aligned}
 \right.,
 \label{eq:2zone}
\end{equation}
where $\mathbi{r}_s$ is source position and $r_\star$ is the size of the 
slow-diffusion zone. The inner diffusion coefficient $D_1$ will be decided by 
fitting the SBP, while the outer value $D_2$ is assumed to be the average value 
in the Galaxy \cite{Yuan:2017ozr}.

The second and third terms on the right-hand side of Eq.~(\ref{eq:prop}) are the energy-loss and source terms, respectively. Synchrotron radiation and ICS dominate the energy losses of high-energy electrons. The magnetic field at the pulsar position should not be very different from the local value considering the radial distribution of the Galactic magnetic field \cite{Moskalenko:1997gh}. We take the local magnetic field strength (3~$\mu$G, \cite{1996ApJ...458..194M}) for the synchrotron component. We adopt the method given in Ref.~\cite{Fang:2020dmi} to get the ICS component, while the seed photon field of ICS is introduced in Sec.~\ref{subsec:en_loss}. The source function $Q$ is introduced in Sec.~\ref{subsec:src}.

For the superdiffusion case, Eq.~(\ref{eq:prop}) can be solved with 
the Green's function method. We directly show the final solution below:
\begin{equation}
 N(E_e, \mathbi{r}, t) = \int_{R^3} d^3\mathbi{r}_0\int_{t_{\rm ini}}^{t}dt_0\,
\frac{b(E_e^\star)}{b(E_e)}\frac{\rho^{(\alpha)}_3(|\mathbi{r}-\mathbi{r}
_0|\lambda^{ -1/\alpha})}{\lambda^{3/\alpha}}\,Q(E_e^\star,
\mathbi{r}_0, t_0)\,,
 \label{eq:solution}
\end{equation}
where
\begin{equation}
 E_e^\star\simeq \frac{E_e}{[1-b_0E_e(t-t_0)]}\,,\quad 
\lambda=\int_{E_e}^{E_e^\star}\frac{D(\alpha,
E'_e)}{b(E'_e)}dE'_e\,,
 \label{eq:cooling}
\end{equation}
and $\rho_3^{(\alpha)}(r)$ is the probability density function of a 
three-dimensional spherically-symmetrical stable distribution with index 
$\alpha$ and expressed as
\begin{equation}
  \rho_3^{(\alpha)}(r) = \frac{1}{2{\pi^2}r} \int_0^\infty{e^{k^{\alpha}}
\sin(kr)kdk}\,.
 \label{eq:rho}
\end{equation}
When $\alpha=2$ or 1, $\rho_3^{(\alpha)}(r)$ is the Gaussian 
distribution or the three-dimensional Cauchy distribution, respectively. 
The lower limit of the time integral is $t_{\rm ini}={\rm max}\{t-1/(b_0E_e), 0\}$.

For the two-zone diffusion case, we adopt the numerical method 
introduced in Ref.~\cite{Fang:2018qco} to solve the propagation equation. The 
finite volume method is used to derive the differencing scheme as there is a 
discontinuity in the diffusion coefficient. One may refer to 
Ref.~\cite{Fang:2018qco} for details.

For both the superdiffusion and two-zone diffusion cases, we integrate $N$ over 
the line of sight from the Earth to the vicinity of the pulsar and get the 
electron surface density:
\begin{equation}
 S_e(\theta)=\int_0^{\infty}N(l_\theta)dl_\theta\,,
 \label{eq:los}
\end{equation}
where $\theta$ is the angle observed away from the pulsar, $l_\theta$ is the 
length in that direction, and $N(l_\theta)$ is the electron number density 
at a distance of $\sqrt{d^2+l_\theta^2-2dl_\theta\cos\theta}$ from the pulsar, 
where $d$ is the distance between the pulsar and the Earth. 

\subsection{Source function}
\label{subsec:src}
The information of PSR J0622$+$3749 can be found in the Australia Telescope 
National Facility catalog \cite{Manchester:2004bp}. The pulsar age and current 
spin-down luminosity are $t_s=208$ kyr and $L=2.7\times10^{34}$ erg s$^{-1}$, 
respectively. The pulsar distance is 1.6 kpc, which is derived from the 
correlation between the gamma-ray luminosity and spin-down luminosity of 
gamma-ray pulsars \cite{Abdo:2009ax}. The electrons are injected from the PWN, 
while the assumed PWN is currently not observed in radio or x-ray bands. It may 
be due to the relatively large distance of the pulsar as discussed in 
Ref.~\cite{Aharonian:2021jtz}. Considering the pulsar age and the evolution 
model of PWN \cite{Gaensler:2006ua}, the PWN should be much smaller than the 
TeV halo and we can safely assume it to be a point-like source. The time 
dependency of the electron injection is assumed to be proportional to the 
spin-down luminosity of the pulsar as $\propto(1+t/t_{\rm sd})^{-2}$, where the 
spin-down time scale is set to be $t_{\rm sd}=10$ kyr. Hence, the source 
function is expressed as 
\begin{equation}
 Q(E_e,\mathbi{r},t)=\left\{
 \begin{aligned}
 & q(E_e)\,\delta(\mathbi{r}-\mathbi{r}_s)\,[(t_s+t_ { \rm
sd})/(t+t_{\rm sd})]^2\,, & t\geq0 \\
 & 0\,, & t<0
 \end{aligned}
 \right.\,,
 \label{eq:src}
\end{equation}
where $q(E_e)$ is the electron injection spectrum.

To simultaneously explain the low-energy Fermi-LAT ULs and the high-energy 
LHAASO-KM2A data, the injection spectrum could be a power-law form with a 
high-energy cutoff: 
\begin{equation}
 q(E_e)=q_0E_e^{-p}\,{\rm exp}[-(E_e/E_c)^2]\,,
 \label{eq:inj}
\end{equation}
where the super-exponential cutoff term is suggested for the spectrum of 
shock-accelerated electrons \cite{Zirakashvili:2006pv}. The power-law spectral 
index may be estimated from the observations of other 
PWNe. Since the electron energy corresponding to the x-ray synchrotron emission 
may be close to $E_c$, the radio spectral indices of PWNe could be the more 
proper indicators. The average electron spectral index of observed radio PWNe 
is $\sim1.5$ \cite{2017SSRv..207..175R}, and we set $p=1.5$ as default. The 
energy spectrum is related with the spin-down luminosity $L$ by 
\begin{equation}
 \int q(E_e)E_edE_e=\eta L\,,
 \label{eq:eta}
\end{equation}
where $\eta$ is the conversion efficiency from the spin-down energy to the 
electron energy. When $E_c$ and $p$ are determined, there is a one-to-one 
correspondence between $\eta$ and $q_0$. Since the physical meaning of $\eta$ 
is more explicit, we choose it as the fitting parameter instead of $q_0$ in the 
following sections. When $p<2.0$, the energy of the electron spectrum is 
concentrated around the cutoff energy, and the LHAASO-KM2A data can well constrain $\eta$.

\subsection{Seed photon field of ICS}
\label{subsec:en_loss}
The seed photon field of ICS consists of the cosmic microwave background (CMB), the infrared dust emission, and the starlight. The temperature and energy density of CMB are 2.725 K and 0.26 eV cm$^{-3}$ \cite{Fixsen:2009ug}. We adopt the methods introduced in Ref.~\cite{Vernetto:2016alq} to get the infrared and starlight components; the infrared component is more important for the energy range we are interested in. The energy and space dependencies of the infrared emission are obtained by fitting the spectral and angular distributions of COBE-FIRAS and COBE-DIRBE \cite{Misiriotis:2006qq}. We simplify the infrared and starlight components by searching for the best-fit gray body distributions to them, respectively. Considering the position of PSR J0622$+$3749, the temperatures and energy densities of the infrared and starlight components are respectively 29 K, 0.11 eV cm$^{-3}$ and 4300 K, 0.22 eV cm$^{-3}$. We use this photon field in the calculations of electron energy loss and gamma-ray emission.

\section{Analysis of Fermi-LAT Data}
\label{sec:fermi}
Fermi-LAT is an imaging, wide field of view, pair conversion telescope, covering the energy from $\sim20$~MeV up to $>500$~GeV \cite{Fermi-LAT:2009ihh}. This work uses $\sim$12~years (MET 
239557417-625393779) of the data belonging to the Pass 8 SOURCE event class represented by the P8R3\_SOURCE\_V2 instrument response functions. We employ the Science Tools package (v11r5p3) to perform a binned analysis for Fermi-LAT data. We select photons with energies from 15~GeV to 500~GeV within a $40^\circ \times 40^\circ$ region of interest (ROI) centered on the position of LHAASO J0621+3749 at $\alpha_{2000}$=95.47 and $\delta_{2000}$=37.92. Limiting the data selection to zenith angles less than $105^\circ$ allows us to effectively exclude the contamination of the photons originating from the Earth limb for the analysis above 10~GeV energy. We further use $gtmktime$ tool to select good time intervals defined by expression DATA\_QUAL$>$0\&\&LAT\_CONFIG==1. We bin the data with a pixel size of $0.1^\circ$ and eight bins per energy decade.

The $\gamma$-ray photons in our ROI are contributed by the Galactic diffuse emission and isotropic diffuse emission, as well as the astrophysical sources extended $30^\circ$ from the ROI center. We created our background source model including the diffuse models shaped  by  gll\_iem\_v07.fits and iso\_P8R3\_SOURCE\_V2\_v1.txt, and the point-like and extended sources listed in 4FGL source catalog\cite{Fermi-LAT:2016zaq,Ballet:2020hze}. We find no obvious emission around LHAASO J0621+3749 after subtracting the contribution of the background sources, as reported by Ref.~\cite{Aharonian:2021jtz}. The 95\% flux upper limits are then derived for those spatial template predicted by the diffusion model with a $20^\circ$ cut in the relevant energy band.

\section{Superdiffusion scenario}
\label{sec:super}
We first fit the SBP measured by LHAASO-KM2A to obtain the diffusion coefficients of superdiffusion models with different $\alpha$. The diffusion coefficients are extrapolated to lower energies and used to generate the spatial templates for the Fermi-LAT analysis. Then we compare the theoretical spectra with the multi-wavelength gamma-ray data to test the superdiffusion models.

\subsection{Fit to the TeV gamma-ray morphology}
\label{subsec:prof}
The SBP of the halo is mainly decided by the diffusion coefficient and has a weak dependence on the shape of the injection spectrum. We first determine the electron injection spectrum by fitting the whole-space gamma-ray spectrum given by LHAASO-KM2A, where only the energy-loss process for electrons needs to be considered. The free parameters are $E_c$ and $\eta$. The power-law term of the injection spectrum cannot be constrained by the LHAASO-KM2A data, and we keep the spectral index $p$ as the default value. We use the $\chi^2$-fitting to search the best-fit parameters. The fitting result is 
$E_c=264^{+62}_{-50}$~TeV and $\eta=0.40^{+0.10}_{-0.08}$.

\begin{figure}[t]
\centering
\includegraphics[width=0.48\textwidth]{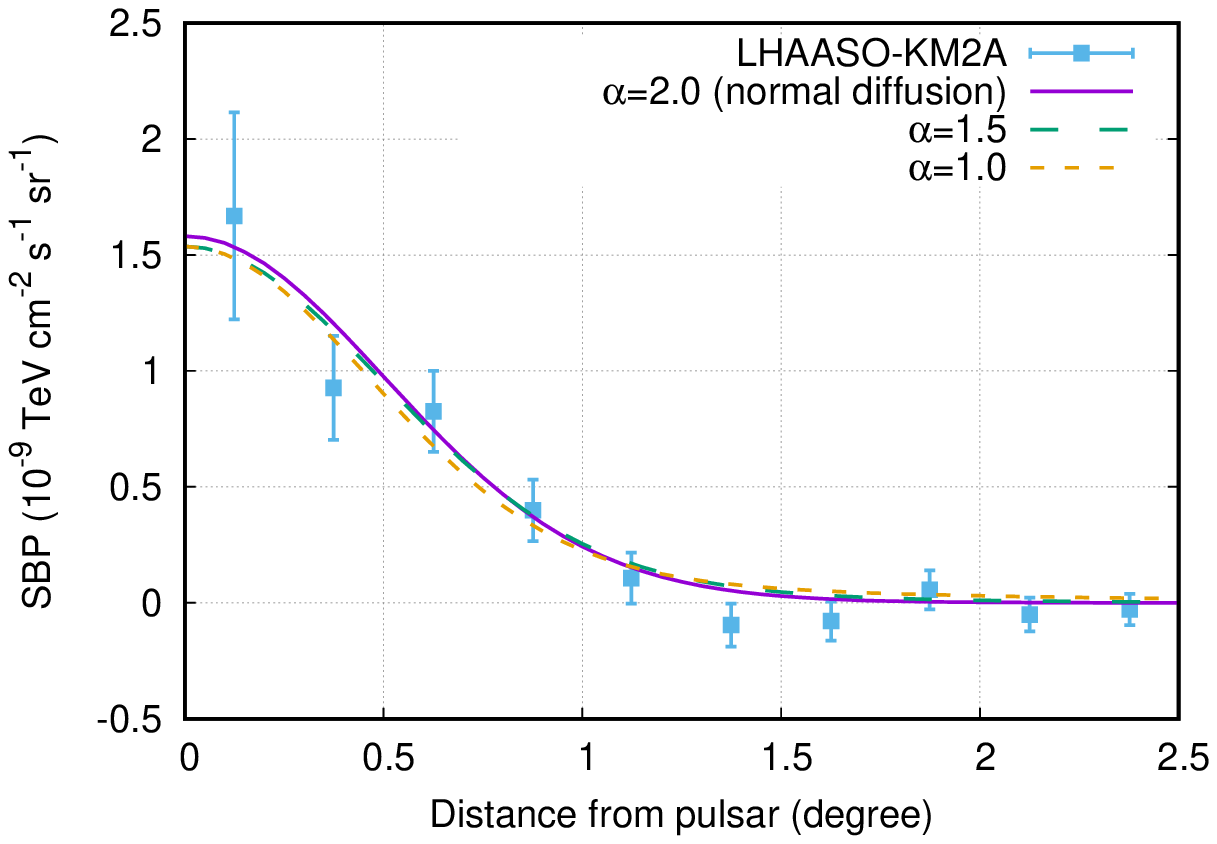}
\includegraphics[width=0.48\textwidth]{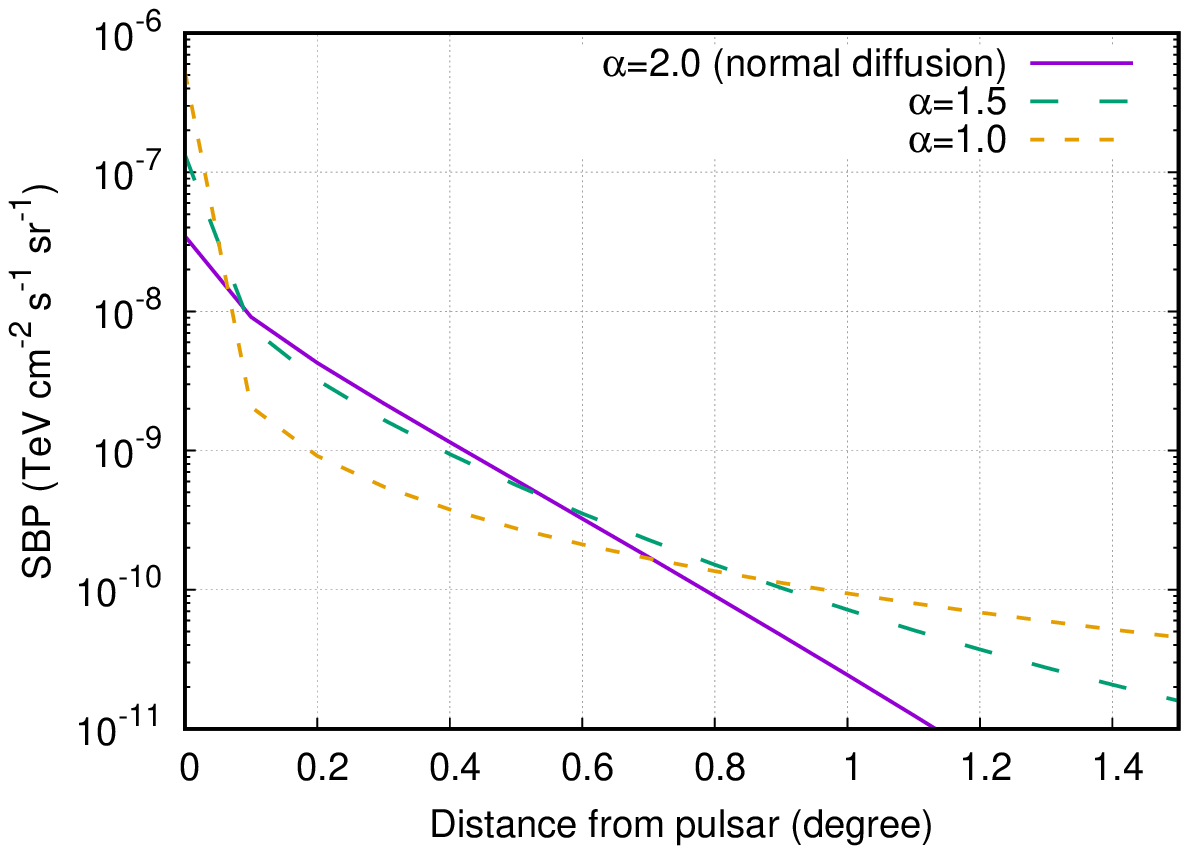}
\caption{Left: best-fit SBPs to the LHAASO-KM2A data with both the normal diffusion ($\alpha=2$) and superdiffusion ($\alpha=1.5$, 1.0) models. Right: SBPs before the convolution with the PSF, corresponding to the results in the left.}
\label{fig:profile}
\end{figure}

Then we fit the SBP with the normal diffusion and superdiffusion models, respectively. The differential surface brightness of gamma rays, $S_\gamma(\theta, E_\gamma)$, is derived from Eq.~(\ref{eq:los}) and the standard calculation of ICS. The flux points of LHAASO-KM2A is the gamma-ray emission above 25 TeV, so we integrate $S_\gamma$ over the gamma-ray energy to match the data, which is written as $\int_{\rm 25 TeV}^\infty S_\gamma(\theta, E_\gamma)E_\gamma dE_\gamma$. As the injection spectrum has been determined, the only free parameter is the diffusion coefficient for each propagation model. Unlike the case of Geminga, the angular extension of the halo is not significantly larger than the width of the point-spread function (PSF). We need to convolve the SBP with the PSF, which is a Gaussian function with a size of 0.45$^\circ$ \cite{Aharonian:2021jtz}.

The best-fit SBPs for three different propagation models ($\alpha=2$, 1.5, and 1) are shown in the left penal of Fig.~\ref{fig:profile}, compared with the LHAASO-KM2A flux points. All the propagation models explain the data well, and the reduced $\chi^2$ statistics are around 1. We also show the SBPs before the convolution with the PSF in the right panel of Fig.~\ref{fig:profile}. The distributions before the convolution are all significantly different, while the distinct features are smoothed by the PSF.

The best-fit diffusion coefficients at 100 TeV for the cases of $\alpha=2$, 1.5, and 1 are $2.5\times10^{27}$ cm$^2$ s$^{-1}$, $8.7\times10^{17}$ cm$^{1.5}$ s$^{-1}$, and $1.1\times10^{9}$ cm s$^{-1}$, respectively. The diffusion coefficient of the normal diffusion model is very similar to that of Geminga, which is $3.2\times10^{27}$ cm$^2$ s$^{-1}$ at 100 TeV as measured by HAWC \cite{Abeysekara:2017old}. Considering the other similarities, the slow-diffusion zone around PSR J0622$+$3749 is very likely to share the same origin with that of Geminga. 

\begin{figure}[t]
\centering
\includegraphics[width=0.68\textwidth]{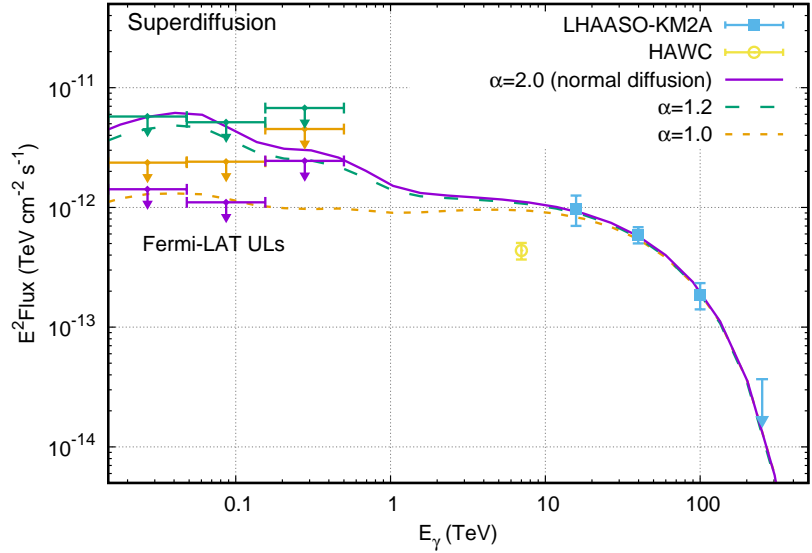}
\caption{Comparison between the gamma-ray spectra calculated with superdiffusion models and the multi-wavelength observations. The theoretical spectra shown here are the integrated fluxes within 20$^\circ$ around the pulsar, and so do the Fermi-LAT ULs. The power-law index of the electron 
injection spectrum is 1.5 for all the cases.}
\label{fig:spec_spr}
\end{figure}

\subsection{Interpretation of the gamma-ray spectrum}
\label{subsec:spec}
Observation in the energy range of Fermi-LAT is important for a comprehensive understanding of the pulsar halo as it can provide information complementary to the measurement of LHAASO-KM2A. Although no significant extended emission is detected by Fermi-LAT around PSR J0622$+$3749, the flux ULs given by Fermi-LAT can be very helpful to test theoretical models. Using the diffusion coefficients extrapolated from the high-energy range, we generate the templates for the observation of Fermi-LAT. As introduced in Sec.~\ref{sec:fermi}, we cut the templates at 20$^\circ$, and the templates are calculated by $\int_{0^\circ}^{20^\circ} S_\gamma(\theta, E_\gamma)2\pi\theta d\theta$.

In Fig.~\ref{fig:spec_spr}, we compare the theoretical models with the multi-wavelength gamma-ray observations. For the normal diffusion case, the predicted GeV spectrum is significantly higher than the corresponding ULs of Fermi-LAT. As indicated by Fig.~S4 of the Supplemental Material of Ref.~\cite{Aharonian:2021jtz}, only an energy-independent injection spectrum, which is unreasonable, can marginally solve this conflict. Since the conflict is significant, it can hardly be explained by adjusting the ISRF or the ambient magnetic field within a reasonable range or assuming an energy-independent diffusion coefficient. Thus, the normal diffusion model is strongly disfavored by the constraint of Fermi-LAT observation.

As shown in Fig.~\ref{fig:spec_spr}, superdiffusion models with $\alpha=1.2$ and 1 can keep the spectra under the Fermi-LAT ULs. Especially, the GeV fluxes predicted by the $\alpha=1$ case are more than two times lower than the ULs. The microscopic particle motion for a superdiffusion model is L\'{e}vy flight instead of the Brownian motion. The individual steps of L\'{e}vy flight are distributed by the heavy-tailed form, which permits extremely long jumps compared with the Brownian motion. As a result, the widening of the diffusion packet with time is proportional to $t^{1/\alpha}$ for a superdiffusion model ($\alpha<2$), faster than the $\propto t^{1/2}$ predicted by the normal diffusion. Consequently, a superdiffusion model with a larger $\alpha$ roughly results in a smaller extension and larger expected fluxes in the $20^\circ$ cut region and thus tends to be constrained by Fermi-LAT observation.

\section{Two-zone diffusion scenario}
\label{sec:2zone}
\begin{table}[t]
 \centering
 \caption{Maximum size of the slow-diffusion zone around PSR J0622$+$3749 with 
varying injection spectral index. The best-fit $E_c$ and $\eta$ for each case 
are also shown.}
 \begin{tabular}{lrrrrr}
  \hline
  \hline
  $p$ & 1.2 & 1.35 & 1.5 & 1.65 & 1.8 \\
  \hline
  $r_{\star, {\rm max}}$ (pc) & \quad$40-50$ & \quad$40-50$ & \quad$30-40$ & 
  \quad$30-40$ & \quad$20-30$ \\
  $E_c$ (TeV) & 232 & 249 & 265 & 284 & 307 \\
  $\eta$ & 0.30 & 0.34 & 0.40 & 0.51 & 0.74 \\
  \hline
 \end{tabular}
 \label{tab:var_p}
\end{table}

We discuss the two-zone diffusion scenario with a process similar to that of the superdiffusion case. For different sizes of the slow-diffusion zone, the fitting results to the SBP are similar to those in Fig.~\ref{fig:profile} and are not shown here. We note that for $r_\star\geq30$~pc, the best-fit $D_1$ is very close to the best-fit diffusion coefficient of the normal diffusion case obtained in Sec.~\ref{subsec:prof}. As most high-energy electrons may still be trapped in the slow-diffusion zone, the electron distribution of a two-zone diffusion model can be similar to that of the normal diffusion case in the inner region \cite{Fang:2018qco}.

\begin{figure}[t]
\centering
\includegraphics[width=0.68\textwidth]{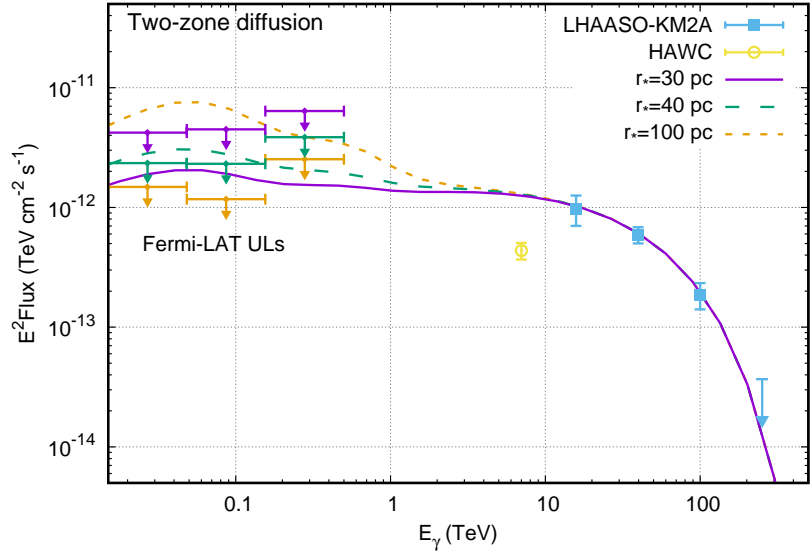}
\caption{Same as Fig.~\ref{fig:spec_spr} but for the two-zone diffusion models.}
\label{fig:spec_2zone}
\end{figure}

We calculate the wide-band gamma-ray spectra for different $r_\star$ and compare the results with the observations in Fig.~\ref{fig:spec_2zone}. The case of $r_\star=30$~pc is obviously permitted by the Fermi-LAT ULs, while the $r_\star=40$~pc case is marginally excluded. We also show that a large slow-diffusion zone with $r_\star=100$~pc is strongly disfavored. The maximum size $r_{\star, {\rm max}}$ of the slow-diffusion zone around the pulsar should be $30-40$~pc for the case of $p=1.5$.

The maximum size of the slow-diffusion zone depends on the injection spectrum. When $p$ is larger, the constraint from Fermi-LAT observation is stronger and $r_{\star, {\rm max}}$ should be smaller, and vice versa. We repeat the above calculations for different $p$ and summarize the results in Table~\ref{tab:var_p}. We find that $p$ cannot be larger than 1.9 or the required conversion efficiency is larger than 100\%. The results indicate that $r_{\star, {\rm max}}$ should not be larger than $\sim50$ pc for a reasonable $p$. This is consistent with the expectation for the self-excited or the SNR-associated origin of the slow-diffusion zone \cite{Evoli:2018aza,Kun:2019sks}.

\begin{figure}[t]
\centering
\includegraphics[width=0.48\textwidth]{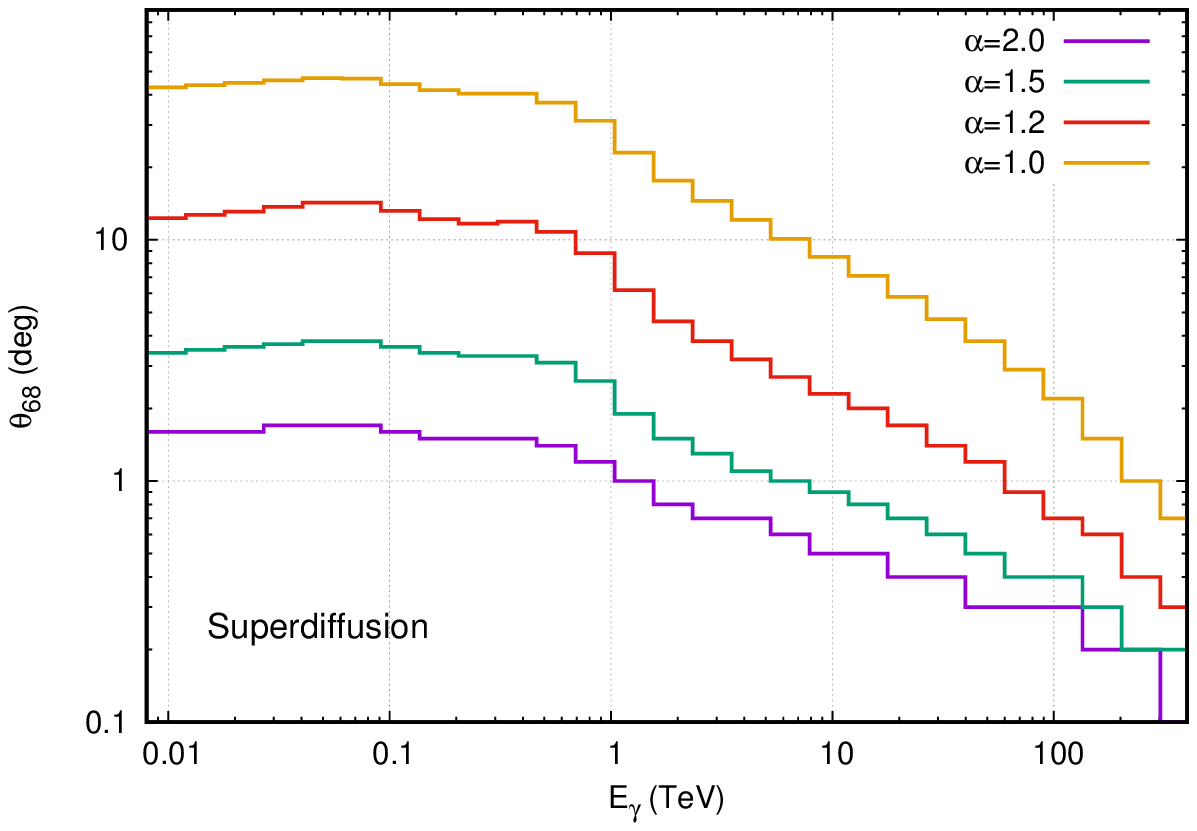}
\includegraphics[width=0.48\textwidth]{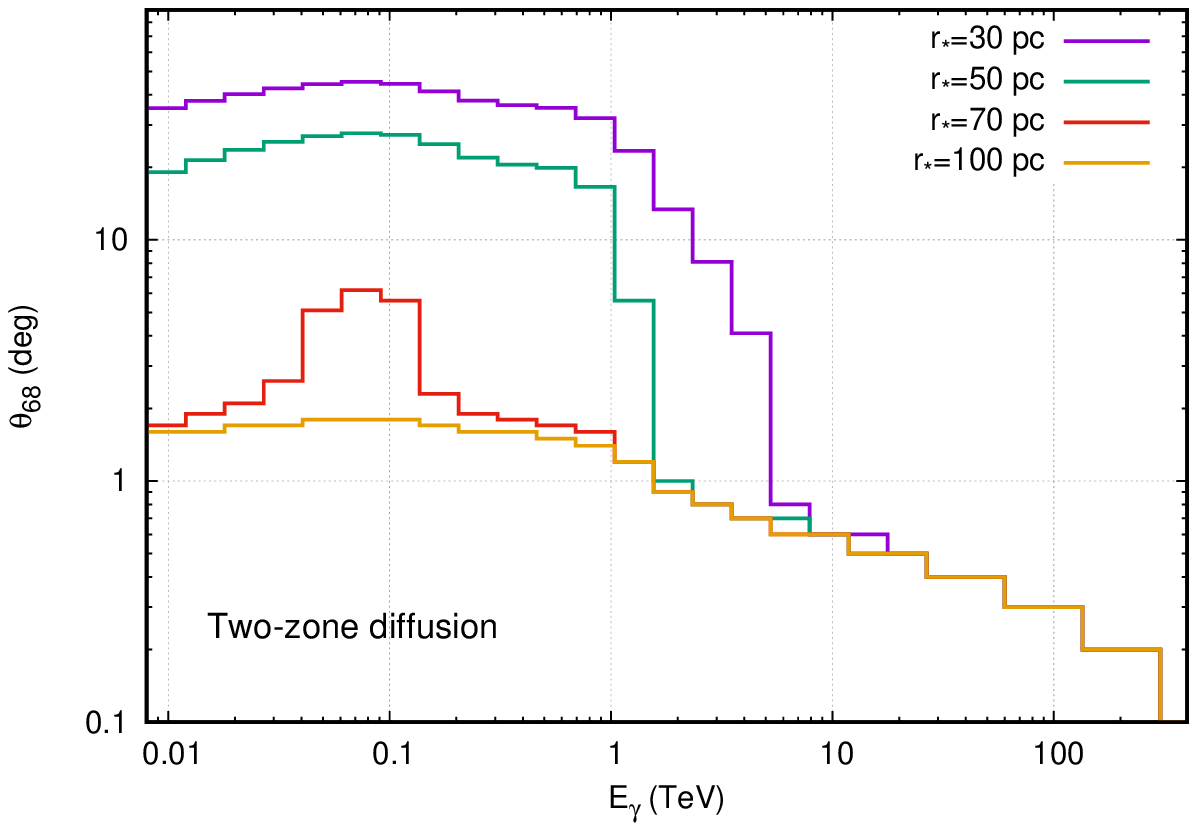}
\caption{Gamma-ray extensions as functions of energy for the superdiffusion 
(left) or the two-zone diffusion (right) scenario. The extension is defined as 
the angular size within which $68\%$ of the gamma-ray flux is included. }
\label{fig:tht68}
\end{figure}

In Fig.~\ref{fig:tht68} we show the gamma-ray extension as a function of energy for both the two-zone diffusion and superdiffusion models. The extension of each model, denoted by $\theta_{68}$, is defined as the angular size within which $68\%$ of the gamma-ray flux is included. This quantity can provide a direct understanding of the calculated spectra. For example, low-energy electrons can significantly escape from the slow-diffusion zone for $r_\star=30$~pc while they are still trapped in the inner zone for the case of $r_\star=100$~pc. Thus, within the 20$^\circ$ cut region, the predicted fluxes below 1~TeV of the former are significantly smaller than that of the latter, as shown in Fig.~\ref{fig:spec_2zone}.

It is worth noting that the flux measured by HAWC is significantly lower than all the theoretical calculations above \cite{Albert:2020fua}. This flux was derived assuming a disk extension of $0.5^\circ$. As shown in Fig.~\ref{fig:tht68}, the extension under the superdiffusion or the two-zone models could be significantly larger than $0.5^\circ$ in the energy of the HAWC measurement. This implies that the whole-space flux may be much higher than the current result of HAWC.

\section{Conclusion}
\label{sec:conclude}
In this work, we simultaneously explain the LHAASO-KM2A and Fermi-LAT observations of the plausible pulsar halo LHAASO~J0621$+$3755 with the superdiffusion and two-zone diffusion models for electron propagation, respectively. The generally used normal diffusion model is seriously constrained by the Fermi-LAT ULs when the LHAASO-KM2A spectrum is fitted. Both the superdiffusion and two-zone diffusion models can predict much larger gamma-ray extensions in GeV bands than the normal diffusion case. The integrated GeV fluxes within the $20^\circ$ cut region can then be lower, and the corresponding Fermi-LAT ULs are found to be higher for these two models. As a result, the GeV fluxes calculated by these models can be consistent with the ULs of Fermi-LAT.

For the superdiffusion scenario, a model with $\alpha$ close to 1 ($\alpha\leq1.2$ for $p=1.5$) can meet the flux constraints of Fermi-LAT. This index describes the fractal feature of the ISM (the superdiffusion degenerates to the normal diffusion when $\alpha=2$). Superdiffusion with $\alpha$ close to 1 could exist in the turbulent magnetic field as indicated by three-dimensional simulations \cite{1995PhPl....2.2653Z}. For the two-zone diffusion scenario, a model with a smaller slow-diffusion zone is more likely to satisfy the constraints of Fermi-LAT. Assuming a reasonable injection spectrum, we find that the slow-diffusion zone should be smaller than $\sim50$ pc, which is consistent with the theoretical expectations \cite{Evoli:2018aza,Kun:2019sks}. This is the first constraint on the size of the slow-diffusion zone related to pulsar halos under the two-zone diffusion assumption. The slow-diffusion size around pulsars is crucial for the pulsar interpretation of the cosmic positron excess \cite{Fang:2019ayz}.

The current observations can hardly distinguish between the superdiffusion and two-zone diffusion scenarios for the case of LHAASO~J0621$+$3755. As mentioned above, the SBP predicted by a two-zone diffusion model can be very similar to that of the normal diffusion model in the inner region, while a superdiffusion model may give a quite different SBP in the inner region due to the nature of L\'{e}vy flight \cite{Wang:2021xph}. However, the different features are smoothed by the PSF as shown in Sec.~\ref{subsec:prof}. In contrast, the Geminga halo has a much larger extension than the PSF due to its close distance to the Earth, and the features of electron propagation may be preserved in the measured SBP. In the coming future, LHAASO will provide a more precise measurement for the SBP of the Geminga halo, which may clarify the electron propagation in pulsar halos.

\section*{Acknowledgement}
This work is supported by the National Key R\&D Program of China (Grant No. 
2016YFA0400200) and the National Natural Science Foundation of China (Grants No. 
U1738209 and No. 11851303).

\bibliography{references}

\begin{thebibliography}{40}%
\makeatletter
\providecommand \@ifxundefined [1]{%
 \@ifx{#1\undefined}
}%
\providecommand \@ifnum [1]{%
 \ifnum #1\expandafter \@firstoftwo
 \else \expandafter \@secondoftwo
 \fi
}%
\providecommand \@ifx [1]{%
 \ifx #1\expandafter \@firstoftwo
 \else \expandafter \@secondoftwo
 \fi
}%
\providecommand \natexlab [1]{#1}%
\providecommand \enquote  [1]{``#1''}%
\providecommand \bibnamefont  [1]{#1}%
\providecommand \bibfnamefont [1]{#1}%
\providecommand \citenamefont [1]{#1}%
\providecommand \href@noop [0]{\@secondoftwo}%
\providecommand \href [0]{\begingroup \@sanitize@url \@href}%
\providecommand \@href[1]{\@@startlink{#1}\@@href}%
\providecommand \@@href[1]{\endgroup#1\@@endlink}%
\providecommand \@sanitize@url [0]{\catcode `\\12\catcode `\$12\catcode
  `\&12\catcode `\#12\catcode `\^12\catcode `\_12\catcode `\%12\relax}%
\providecommand \@@startlink[1]{}%
\providecommand \@@endlink[0]{}%
\providecommand \url  [0]{\begingroup\@sanitize@url \@url }%
\providecommand \@url [1]{\endgroup\@href {#1}{\urlprefix }}%
\providecommand \urlprefix  [0]{URL }%
\providecommand \Eprint [0]{\href }%
\providecommand \doibase [0]{http://dx.doi.org/}%
\providecommand \selectlanguage [0]{\@gobble}%
\providecommand \bibinfo  [0]{\@secondoftwo}%
\providecommand \bibfield  [0]{\@secondoftwo}%
\providecommand \translation [1]{[#1]}%
\providecommand \BibitemOpen [0]{}%
\providecommand \bibitemStop [0]{}%
\providecommand \bibitemNoStop [0]{.\EOS\space}%
\providecommand \EOS [0]{\spacefactor3000\relax}%
\providecommand \BibitemShut  [1]{\csname bibitem#1\endcsname}%
\let\auto@bib@innerbib\@empty
\bibitem [{\citenamefont {Linden}\ \emph {et~al.}(2017)\citenamefont {Linden},
  \citenamefont {Auchettl}, \citenamefont {Bramante}, \citenamefont {Cholis},
  \citenamefont {Fang}, \citenamefont {Hooper}, \citenamefont {Karwal},\ and\
  \citenamefont {Li}}]{Linden:2017vvb}%
  \BibitemOpen
  \bibfield  {author} {\bibinfo {author} {\bibfnamefont {T.}~\bibnamefont
  {Linden}}, \bibinfo {author} {\bibfnamefont {K.}~\bibnamefont {Auchettl}},
  \bibinfo {author} {\bibfnamefont {J.}~\bibnamefont {Bramante}}, \bibinfo
  {author} {\bibfnamefont {I.}~\bibnamefont {Cholis}}, \bibinfo {author}
  {\bibfnamefont {K.}~\bibnamefont {Fang}}, \bibinfo {author} {\bibfnamefont
  {D.}~\bibnamefont {Hooper}}, \bibinfo {author} {\bibfnamefont
  {T.}~\bibnamefont {Karwal}}, \ and\ \bibinfo {author} {\bibfnamefont {S.~W.}\
  \bibnamefont {Li}},\ }\href {\doibase 10.1103/PhysRevD.96.103016} {\bibfield
  {journal} {\bibinfo  {journal} {Phys. Rev. D}\ }\textbf {\bibinfo {volume}
  {96}},\ \bibinfo {pages} {103016} (\bibinfo {year} {2017})},\ \Eprint
  {http://arxiv.org/abs/1703.09704} {arXiv:1703.09704 [astro-ph.HE]}
  \BibitemShut {NoStop}%
\bibitem [{\citenamefont {Sudoh}\ \emph {et~al.}(2019)\citenamefont {Sudoh},
  \citenamefont {Linden},\ and\ \citenamefont {Beacom}}]{Sudoh:2019lav}%
  \BibitemOpen
  \bibfield  {author} {\bibinfo {author} {\bibfnamefont {T.}~\bibnamefont
  {Sudoh}}, \bibinfo {author} {\bibfnamefont {T.}~\bibnamefont {Linden}}, \
  and\ \bibinfo {author} {\bibfnamefont {J.~F.}\ \bibnamefont {Beacom}},\
  }\href {\doibase 10.1103/PhysRevD.100.043016} {\bibfield  {journal} {\bibinfo
   {journal} {Phys. Rev. D}\ }\textbf {\bibinfo {volume} {100}},\ \bibinfo
  {pages} {043016} (\bibinfo {year} {2019})},\ \Eprint
  {http://arxiv.org/abs/1902.08203} {arXiv:1902.08203 [astro-ph.HE]}
  \BibitemShut {NoStop}%
\bibitem [{\citenamefont {Giacinti}\ \emph {et~al.}(2020)\citenamefont
  {Giacinti}, \citenamefont {Mitchell}, \citenamefont {L\'opez-Coto},
  \citenamefont {Joshi}, \citenamefont {Parsons},\ and\ \citenamefont
  {Hinton}}]{Giacinti:2019nbu}%
  \BibitemOpen
  \bibfield  {author} {\bibinfo {author} {\bibfnamefont {G.}~\bibnamefont
  {Giacinti}}, \bibinfo {author} {\bibfnamefont {A.}~\bibnamefont {Mitchell}},
  \bibinfo {author} {\bibfnamefont {R.}~\bibnamefont {L\'opez-Coto}}, \bibinfo
  {author} {\bibfnamefont {V.}~\bibnamefont {Joshi}}, \bibinfo {author}
  {\bibfnamefont {R.}~\bibnamefont {Parsons}}, \ and\ \bibinfo {author}
  {\bibfnamefont {J.}~\bibnamefont {Hinton}},\ }\href {\doibase
  10.1051/0004-6361/201936505} {\bibfield  {journal} {\bibinfo  {journal}
  {Astron. Astrophys.}\ }\textbf {\bibinfo {volume} {636}},\ \bibinfo {pages}
  {A113} (\bibinfo {year} {2020})},\ \Eprint {http://arxiv.org/abs/1907.12121}
  {arXiv:1907.12121 [astro-ph.HE]} \BibitemShut {NoStop}%
\bibitem [{\citenamefont {Abeysekara}\ \emph {et~al.}(2017)\citenamefont
  {Abeysekara} \emph {et~al.}}]{Abeysekara:2017old}%
  \BibitemOpen
  \bibfield  {author} {\bibinfo {author} {\bibfnamefont {A.}~\bibnamefont
  {Abeysekara}} \emph {et~al.} (\bibinfo {collaboration} {HAWC}),\ }\href
  {\doibase 10.1126/science.aan4880} {\bibfield  {journal} {\bibinfo  {journal}
  {Science}\ }\textbf {\bibinfo {volume} {358}},\ \bibinfo {pages} {911}
  (\bibinfo {year} {2017})},\ \Eprint {http://arxiv.org/abs/1711.06223}
  {arXiv:1711.06223 [astro-ph.HE]} \BibitemShut {NoStop}%
\bibitem [{\citenamefont {Evoli}\ \emph {et~al.}(2018)\citenamefont {Evoli},
  \citenamefont {Linden},\ and\ \citenamefont {Morlino}}]{Evoli:2018aza}%
  \BibitemOpen
  \bibfield  {author} {\bibinfo {author} {\bibfnamefont {C.}~\bibnamefont
  {Evoli}}, \bibinfo {author} {\bibfnamefont {T.}~\bibnamefont {Linden}}, \
  and\ \bibinfo {author} {\bibfnamefont {G.}~\bibnamefont {Morlino}},\ }\href
  {\doibase 10.1103/PhysRevD.98.063017} {\bibfield  {journal} {\bibinfo
  {journal} {Phys. Rev. D}\ }\textbf {\bibinfo {volume} {98}},\ \bibinfo
  {pages} {063017} (\bibinfo {year} {2018})},\ \Eprint
  {http://arxiv.org/abs/1807.09263} {arXiv:1807.09263 [astro-ph.HE]}
  \BibitemShut {NoStop}%
\bibitem [{\citenamefont {Fang}\ \emph
  {et~al.}(2019{\natexlab{a}})\citenamefont {Fang}, \citenamefont {Bi},\ and\
  \citenamefont {Yin}}]{Kun:2019sks}%
  \BibitemOpen
  \bibfield  {author} {\bibinfo {author} {\bibfnamefont {K.}~\bibnamefont
  {Fang}}, \bibinfo {author} {\bibfnamefont {X.-J.}\ \bibnamefont {Bi}}, \ and\
  \bibinfo {author} {\bibfnamefont {P.-F.}\ \bibnamefont {Yin}},\ }\href
  {\doibase 10.1093/mnras/stz1974} {\bibfield  {journal} {\bibinfo  {journal}
  {Mon. Not. Roy. Astron. Soc.}\ }\textbf {\bibinfo {volume} {488}},\ \bibinfo
  {pages} {4074} (\bibinfo {year} {2019}{\natexlab{a}})},\ \Eprint
  {http://arxiv.org/abs/1903.06421} {arXiv:1903.06421 [astro-ph.HE]}
  \BibitemShut {NoStop}%
\bibitem [{\citenamefont {Liu}\ \emph {et~al.}(2019)\citenamefont {Liu},
  \citenamefont {Yan},\ and\ \citenamefont {Zhang}}]{Liu:2019zyj}%
  \BibitemOpen
  \bibfield  {author} {\bibinfo {author} {\bibfnamefont {R.-Y.}\ \bibnamefont
  {Liu}}, \bibinfo {author} {\bibfnamefont {H.}~\bibnamefont {Yan}}, \ and\
  \bibinfo {author} {\bibfnamefont {H.}~\bibnamefont {Zhang}},\ }\href
  {\doibase 10.1103/PhysRevLett.123.221103} {\bibfield  {journal} {\bibinfo
  {journal} {Phys. Rev. Lett.}\ }\textbf {\bibinfo {volume} {123}},\ \bibinfo
  {pages} {221103} (\bibinfo {year} {2019})},\ \Eprint
  {http://arxiv.org/abs/1904.11536} {arXiv:1904.11536 [astro-ph.HE]}
  \BibitemShut {NoStop}%
\bibitem [{\citenamefont {Wang}\ \emph {et~al.}(2021)\citenamefont {Wang},
  \citenamefont {Fang}, \citenamefont {Bi},\ and\ \citenamefont
  {Yin}}]{Wang:2021xph}%
  \BibitemOpen
  \bibfield  {author} {\bibinfo {author} {\bibfnamefont {S.-H.}\ \bibnamefont
  {Wang}}, \bibinfo {author} {\bibfnamefont {K.}~\bibnamefont {Fang}}, \bibinfo
  {author} {\bibfnamefont {X.-J.}\ \bibnamefont {Bi}}, \ and\ \bibinfo {author}
  {\bibfnamefont {P.-F.}\ \bibnamefont {Yin}},\ }\href {\doibase
  10.1103/PhysRevD.103.063035} {\bibfield  {journal} {\bibinfo  {journal}
  {Phys. Rev. D}\ }\textbf {\bibinfo {volume} {103}},\ \bibinfo {pages}
  {063035} (\bibinfo {year} {2021})},\ \Eprint
  {http://arxiv.org/abs/2101.01438} {arXiv:2101.01438 [astro-ph.HE]}
  \BibitemShut {NoStop}%
\bibitem [{\citenamefont {Recchia}\ \emph {et~al.}(2021)\citenamefont
  {Recchia}, \citenamefont {Di~Mauro}, \citenamefont {Aharonian}, \citenamefont
  {Donato}, \citenamefont {Gabici},\ and\ \citenamefont
  {Manconi}}]{Recchia:2021kty}%
  \BibitemOpen
  \bibfield  {author} {\bibinfo {author} {\bibfnamefont {S.}~\bibnamefont
  {Recchia}}, \bibinfo {author} {\bibfnamefont {M.}~\bibnamefont {Di~Mauro}},
  \bibinfo {author} {\bibfnamefont {F.~A.}\ \bibnamefont {Aharonian}}, \bibinfo
  {author} {\bibfnamefont {F.}~\bibnamefont {Donato}}, \bibinfo {author}
  {\bibfnamefont {S.}~\bibnamefont {Gabici}}, \ and\ \bibinfo {author}
  {\bibfnamefont {S.}~\bibnamefont {Manconi}},\ }\href@noop {} {\  (\bibinfo
  {year} {2021})},\ \Eprint {http://arxiv.org/abs/2106.02275} {arXiv:2106.02275
  [astro-ph.HE]} \BibitemShut {NoStop}%
\bibitem [{\citenamefont {Hooper}\ \emph {et~al.}(2017)\citenamefont {Hooper},
  \citenamefont {Cholis}, \citenamefont {Linden},\ and\ \citenamefont
  {Fang}}]{Hooper:2017gtd}%
  \BibitemOpen
  \bibfield  {author} {\bibinfo {author} {\bibfnamefont {D.}~\bibnamefont
  {Hooper}}, \bibinfo {author} {\bibfnamefont {I.}~\bibnamefont {Cholis}},
  \bibinfo {author} {\bibfnamefont {T.}~\bibnamefont {Linden}}, \ and\ \bibinfo
  {author} {\bibfnamefont {K.}~\bibnamefont {Fang}},\ }\href {\doibase
  10.1103/PhysRevD.96.103013} {\bibfield  {journal} {\bibinfo  {journal} {Phys.
  Rev. D}\ }\textbf {\bibinfo {volume} {96}},\ \bibinfo {pages} {103013}
  (\bibinfo {year} {2017})},\ \Eprint {http://arxiv.org/abs/1702.08436}
  {arXiv:1702.08436 [astro-ph.HE]} \BibitemShut {NoStop}%
\bibitem [{\citenamefont {Fang}\ \emph {et~al.}(2018)\citenamefont {Fang},
  \citenamefont {Bi}, \citenamefont {Yin},\ and\ \citenamefont
  {Yuan}}]{Fang:2018qco}%
  \BibitemOpen
  \bibfield  {author} {\bibinfo {author} {\bibfnamefont {K.}~\bibnamefont
  {Fang}}, \bibinfo {author} {\bibfnamefont {X.-J.}\ \bibnamefont {Bi}},
  \bibinfo {author} {\bibfnamefont {P.-F.}\ \bibnamefont {Yin}}, \ and\
  \bibinfo {author} {\bibfnamefont {Q.}~\bibnamefont {Yuan}},\ }\href {\doibase
  10.3847/1538-4357/aad092} {\bibfield  {journal} {\bibinfo  {journal}
  {Astrophys. J.}\ }\textbf {\bibinfo {volume} {863}},\ \bibinfo {pages} {30}
  (\bibinfo {year} {2018})},\ \Eprint {http://arxiv.org/abs/1803.02640}
  {arXiv:1803.02640 [astro-ph.HE]} \BibitemShut {NoStop}%
\bibitem [{\citenamefont {Profumo}\ \emph {et~al.}(2018)\citenamefont
  {Profumo}, \citenamefont {Reynoso-Cordova}, \citenamefont {Kaaz},\ and\
  \citenamefont {Silverman}}]{Profumo:2018fmz}%
  \BibitemOpen
  \bibfield  {author} {\bibinfo {author} {\bibfnamefont {S.}~\bibnamefont
  {Profumo}}, \bibinfo {author} {\bibfnamefont {J.}~\bibnamefont
  {Reynoso-Cordova}}, \bibinfo {author} {\bibfnamefont {N.}~\bibnamefont
  {Kaaz}}, \ and\ \bibinfo {author} {\bibfnamefont {M.}~\bibnamefont
  {Silverman}},\ }\href {\doibase 10.1103/PhysRevD.97.123008} {\bibfield
  {journal} {\bibinfo  {journal} {Phys. Rev. D}\ }\textbf {\bibinfo {volume}
  {97}},\ \bibinfo {pages} {123008} (\bibinfo {year} {2018})},\ \Eprint
  {http://arxiv.org/abs/1803.09731} {arXiv:1803.09731 [astro-ph.HE]}
  \BibitemShut {NoStop}%
\bibitem [{\citenamefont {Tang}\ and\ \citenamefont
  {Piran}(2019)}]{Tang:2018wyr}%
  \BibitemOpen
  \bibfield  {author} {\bibinfo {author} {\bibfnamefont {X.}~\bibnamefont
  {Tang}}\ and\ \bibinfo {author} {\bibfnamefont {T.}~\bibnamefont {Piran}},\
  }\href {\doibase 10.1093/mnras/stz268} {\bibfield  {journal} {\bibinfo
  {journal} {Mon. Not. Roy. Astron. Soc.}\ }\textbf {\bibinfo {volume} {484}},\
  \bibinfo {pages} {3491} (\bibinfo {year} {2019})},\ \Eprint
  {http://arxiv.org/abs/1808.02445} {arXiv:1808.02445 [astro-ph.HE]}
  \BibitemShut {NoStop}%
\bibitem [{\citenamefont {Xi}\ \emph {et~al.}(2019)\citenamefont {Xi},
  \citenamefont {Liu}, \citenamefont {Huang}, \citenamefont {Fang},\ and\
  \citenamefont {Wang}}]{Shao-Qiang:2018zla}%
  \BibitemOpen
  \bibfield  {author} {\bibinfo {author} {\bibfnamefont {S.-Q.}\ \bibnamefont
  {Xi}}, \bibinfo {author} {\bibfnamefont {R.-Y.}\ \bibnamefont {Liu}},
  \bibinfo {author} {\bibfnamefont {Z.-Q.}\ \bibnamefont {Huang}}, \bibinfo
  {author} {\bibfnamefont {K.}~\bibnamefont {Fang}}, \ and\ \bibinfo {author}
  {\bibfnamefont {X.-Y.}\ \bibnamefont {Wang}},\ }\href {\doibase
  10.3847/1538-4357/ab20c9} {\bibfield  {journal} {\bibinfo  {journal}
  {Astrophys. J.}\ }\textbf {\bibinfo {volume} {878}},\ \bibinfo {pages} {104}
  (\bibinfo {year} {2019})},\ \Eprint {http://arxiv.org/abs/1810.10928}
  {arXiv:1810.10928 [astro-ph.HE]} \BibitemShut {NoStop}%
\bibitem [{\citenamefont {Di~Mauro}\ \emph {et~al.}(2019)\citenamefont
  {Di~Mauro}, \citenamefont {Manconi},\ and\ \citenamefont
  {Donato}}]{DiMauro:2019yvh}%
  \BibitemOpen
  \bibfield  {author} {\bibinfo {author} {\bibfnamefont {M.}~\bibnamefont
  {Di~Mauro}}, \bibinfo {author} {\bibfnamefont {S.}~\bibnamefont {Manconi}}, \
  and\ \bibinfo {author} {\bibfnamefont {F.}~\bibnamefont {Donato}},\ }\href
  {\doibase 10.1103/PhysRevD.100.123015} {\bibfield  {journal} {\bibinfo
  {journal} {Phys. Rev. D}\ }\textbf {\bibinfo {volume} {100}},\ \bibinfo
  {pages} {123015} (\bibinfo {year} {2019})},\ \Eprint
  {http://arxiv.org/abs/1903.05647} {arXiv:1903.05647 [astro-ph.HE]}
  \BibitemShut {NoStop}%
\bibitem [{\citenamefont {Fang}\ \emph
  {et~al.}(2019{\natexlab{b}})\citenamefont {Fang}, \citenamefont {Bi},\ and\
  \citenamefont {Yin}}]{Fang:2019ayz}%
  \BibitemOpen
  \bibfield  {author} {\bibinfo {author} {\bibfnamefont {K.}~\bibnamefont
  {Fang}}, \bibinfo {author} {\bibfnamefont {X.-J.}\ \bibnamefont {Bi}}, \ and\
  \bibinfo {author} {\bibfnamefont {P.-F.}\ \bibnamefont {Yin}},\ }\href
  {\doibase 10.3847/1538-4357/ab3fac} {\bibfield  {journal} {\bibinfo
  {journal} {Astrophys. J.}\ }\textbf {\bibinfo {volume} {884}},\ \bibinfo
  {pages} {124} (\bibinfo {year} {2019}{\natexlab{b}})},\ \Eprint
  {http://arxiv.org/abs/1906.08542} {arXiv:1906.08542 [astro-ph.HE]}
  \BibitemShut {NoStop}%
\bibitem [{\citenamefont {Aharonian}\ \emph {et~al.}(2021)\citenamefont
  {Aharonian} \emph {et~al.}}]{Aharonian:2021jtz}%
  \BibitemOpen
  \bibfield  {author} {\bibinfo {author} {\bibfnamefont {F.}~\bibnamefont
  {Aharonian}} \emph {et~al.} (\bibinfo {collaboration} {LHAASO}),\ }\href
  {\doibase 10.1103/PhysRevLett.126.241103} {\bibfield  {journal} {\bibinfo
  {journal} {Phys. Rev. Lett.}\ }\textbf {\bibinfo {volume} {126}},\ \bibinfo
  {pages} {241103} (\bibinfo {year} {2021})},\ \Eprint
  {http://arxiv.org/abs/2106.09396} {arXiv:2106.09396 [astro-ph.HE]}
  \BibitemShut {NoStop}%
\bibitem [{\citenamefont {{Veltri}}\ \emph {et~al.}(1998)\citenamefont
  {{Veltri}}, \citenamefont {{Zimbardo}},\ and\ \citenamefont
  {{Pommois}}}]{1998AdSpR..22...55V}%
  \BibitemOpen
  \bibfield  {author} {\bibinfo {author} {\bibfnamefont {P.}~\bibnamefont
  {{Veltri}}}, \bibinfo {author} {\bibfnamefont {G.}~\bibnamefont
  {{Zimbardo}}}, \ and\ \bibinfo {author} {\bibfnamefont {P.}~\bibnamefont
  {{Pommois}}},\ }\href {\doibase 10.1016/S0273-1177(97)01099-5} {\bibfield
  {journal} {\bibinfo  {journal} {Advances in Space Research}\ }\textbf
  {\bibinfo {volume} {22}},\ \bibinfo {pages} {55} (\bibinfo {year}
  {1998})}\BibitemShut {NoStop}%
\bibitem [{\citenamefont {{Lagutin}}\ \emph {et~al.}(2001)\citenamefont
  {{Lagutin}}, \citenamefont {{Nikulin}},\ and\ \citenamefont
  {{Uchaikin}}}]{2001NuPhS..97..267L}%
  \BibitemOpen
  \bibfield  {author} {\bibinfo {author} {\bibfnamefont {A.~A.}\ \bibnamefont
  {{Lagutin}}}, \bibinfo {author} {\bibfnamefont {Y.~A.}\ \bibnamefont
  {{Nikulin}}}, \ and\ \bibinfo {author} {\bibfnamefont {V.~V.}\ \bibnamefont
  {{Uchaikin}}},\ }\href {\doibase 10.1016/S0920-5632(01)01280-4} {\bibfield
  {journal} {\bibinfo  {journal} {Nuclear Physics B Proceedings Supplements}\
  }\textbf {\bibinfo {volume} {97}},\ \bibinfo {pages} {267} (\bibinfo {year}
  {2001})}\BibitemShut {NoStop}%
\bibitem [{\citenamefont {{Volkov}}\ \emph {et~al.}(2015)\citenamefont
  {{Volkov}}, \citenamefont {{Lagutin}},\ and\ \citenamefont
  {{Tyumentsev}}}]{2015JPhCS.632a2027V}%
  \BibitemOpen
  \bibfield  {author} {\bibinfo {author} {\bibfnamefont {N.}~\bibnamefont
  {{Volkov}}}, \bibinfo {author} {\bibfnamefont {A.}~\bibnamefont {{Lagutin}}},
  \ and\ \bibinfo {author} {\bibfnamefont {A.}~\bibnamefont {{Tyumentsev}}},\
  }in\ \href {\doibase 10.1088/1742-6596/632/1/012027} {\emph {\bibinfo
  {booktitle} {Journal of Physics Conference Series}}},\ \bibinfo {series}
  {Journal of Physics Conference Series}, Vol.\ \bibinfo {volume} {632}\
  (\bibinfo {year} {2015})\ p.\ \bibinfo {pages} {012027},\ \Eprint
  {http://arxiv.org/abs/1905.06674} {arXiv:1905.06674 [astro-ph.HE]}
  \BibitemShut {NoStop}%
\bibitem [{\citenamefont {{Perri}}\ \emph {et~al.}(2016)\citenamefont
  {{Perri}}, \citenamefont {{Amato}},\ and\ \citenamefont
  {{Zimbardo}}}]{2016A&A...596A..34P}%
  \BibitemOpen
  \bibfield  {author} {\bibinfo {author} {\bibfnamefont {S.}~\bibnamefont
  {{Perri}}}, \bibinfo {author} {\bibfnamefont {E.}~\bibnamefont {{Amato}}}, \
  and\ \bibinfo {author} {\bibfnamefont {G.}~\bibnamefont {{Zimbardo}}},\
  }\href {\doibase 10.1051/0004-6361/201628767} {\bibfield  {journal} {\bibinfo
   {journal} {\aap}\ }\textbf {\bibinfo {volume} {596}},\ \bibinfo {eid} {A34}
  (\bibinfo {year} {2016})}\BibitemShut {NoStop}%
\bibitem [{\citenamefont {{Zimbardo}}\ and\ \citenamefont
  {{Perri}}(2018)}]{2018MNRAS.478.4922Z}%
  \BibitemOpen
  \bibfield  {author} {\bibinfo {author} {\bibfnamefont {G.}~\bibnamefont
  {{Zimbardo}}}\ and\ \bibinfo {author} {\bibfnamefont {S.}~\bibnamefont
  {{Perri}}},\ }\href {\doibase 10.1093/mnras/sty1438} {\bibfield  {journal}
  {\bibinfo  {journal} {\mnras}\ }\textbf {\bibinfo {volume} {478}},\ \bibinfo
  {pages} {4922} (\bibinfo {year} {2018})}\BibitemShut {NoStop}%
\bibitem [{\citenamefont {Blumenthal}\ and\ \citenamefont
  {Gould}(1970)}]{Blumenthal:1970gc}%
  \BibitemOpen
  \bibfield  {author} {\bibinfo {author} {\bibfnamefont {G.}~\bibnamefont
  {Blumenthal}}\ and\ \bibinfo {author} {\bibfnamefont {R.}~\bibnamefont
  {Gould}},\ }\href {\doibase 10.1103/RevModPhys.42.237} {\bibfield  {journal}
  {\bibinfo  {journal} {Rev. Mod. Phys.}\ }\textbf {\bibinfo {volume} {42}},\
  \bibinfo {pages} {237} (\bibinfo {year} {1970})}\BibitemShut {NoStop}%
\bibitem [{\citenamefont {Yuan}\ \emph {et~al.}(2017)\citenamefont {Yuan},
  \citenamefont {Lin}, \citenamefont {Fang},\ and\ \citenamefont
  {Bi}}]{Yuan:2017ozr}%
  \BibitemOpen
  \bibfield  {author} {\bibinfo {author} {\bibfnamefont {Q.}~\bibnamefont
  {Yuan}}, \bibinfo {author} {\bibfnamefont {S.-J.}\ \bibnamefont {Lin}},
  \bibinfo {author} {\bibfnamefont {K.}~\bibnamefont {Fang}}, \ and\ \bibinfo
  {author} {\bibfnamefont {X.-J.}\ \bibnamefont {Bi}},\ }\href {\doibase
  10.1103/PhysRevD.95.083007} {\bibfield  {journal} {\bibinfo  {journal} {Phys.
  Rev. D}\ }\textbf {\bibinfo {volume} {95}},\ \bibinfo {pages} {083007}
  (\bibinfo {year} {2017})},\ \Eprint {http://arxiv.org/abs/1701.06149}
  {arXiv:1701.06149 [astro-ph.HE]} \BibitemShut {NoStop}%
\bibitem [{\citenamefont {Moskalenko}\ and\ \citenamefont
  {Strong}(1998)}]{Moskalenko:1997gh}%
  \BibitemOpen
  \bibfield  {author} {\bibinfo {author} {\bibfnamefont {I.~V.}\ \bibnamefont
  {Moskalenko}}\ and\ \bibinfo {author} {\bibfnamefont {A.~W.}\ \bibnamefont
  {Strong}},\ }\href {\doibase 10.1086/305152} {\bibfield  {journal} {\bibinfo
  {journal} {Astrophys. J.}\ }\textbf {\bibinfo {volume} {493}},\ \bibinfo
  {pages} {694} (\bibinfo {year} {1998})},\ \Eprint
  {http://arxiv.org/abs/astro-ph/9710124} {arXiv:astro-ph/9710124} \BibitemShut
  {NoStop}%
\bibitem [{\citenamefont {{Minter}}\ and\ \citenamefont
  {{Spangler}}(1996)}]{1996ApJ...458..194M}%
  \BibitemOpen
  \bibfield  {author} {\bibinfo {author} {\bibfnamefont {A.~H.}\ \bibnamefont
  {{Minter}}}\ and\ \bibinfo {author} {\bibfnamefont {S.~R.}\ \bibnamefont
  {{Spangler}}},\ }\href {\doibase 10.1086/176803} {\bibfield  {journal}
  {\bibinfo  {journal} {\apj}\ }\textbf {\bibinfo {volume} {458}},\ \bibinfo
  {pages} {194} (\bibinfo {year} {1996})}\BibitemShut {NoStop}%
\bibitem [{\citenamefont {Fang}\ \emph {et~al.}(2021)\citenamefont {Fang},
  \citenamefont {Bi}, \citenamefont {Lin},\ and\ \citenamefont
  {Yuan}}]{Fang:2020dmi}%
  \BibitemOpen
  \bibfield  {author} {\bibinfo {author} {\bibfnamefont {K.}~\bibnamefont
  {Fang}}, \bibinfo {author} {\bibfnamefont {X.-J.}\ \bibnamefont {Bi}},
  \bibinfo {author} {\bibfnamefont {S.-J.}\ \bibnamefont {Lin}}, \ and\
  \bibinfo {author} {\bibfnamefont {Q.}~\bibnamefont {Yuan}},\ }\href {\doibase
  10.1088/0256-307X/38/3/039801} {\bibfield  {journal} {\bibinfo  {journal}
  {Chin. Phys. Lett.}\ }\textbf {\bibinfo {volume} {38}},\ \bibinfo {pages}
  {039801} (\bibinfo {year} {2021})},\ \Eprint
  {http://arxiv.org/abs/2007.15601} {arXiv:2007.15601 [astro-ph.HE]}
  \BibitemShut {NoStop}%
\bibitem [{\citenamefont {Manchester}\ \emph {et~al.}(2005)\citenamefont
  {Manchester}, \citenamefont {Hobbs}, \citenamefont {Teoh},\ and\
  \citenamefont {Hobbs}}]{Manchester:2004bp}%
  \BibitemOpen
  \bibfield  {author} {\bibinfo {author} {\bibfnamefont {R.~N.}\ \bibnamefont
  {Manchester}}, \bibinfo {author} {\bibfnamefont {G.~B.}\ \bibnamefont
  {Hobbs}}, \bibinfo {author} {\bibfnamefont {A.}~\bibnamefont {Teoh}}, \ and\
  \bibinfo {author} {\bibfnamefont {M.}~\bibnamefont {Hobbs}},\ }\href
  {\doibase 10.1086/428488} {\bibfield  {journal} {\bibinfo  {journal} {Astron.
  J.}\ }\textbf {\bibinfo {volume} {129}},\ \bibinfo {pages} {1993} (\bibinfo
  {year} {2005})},\ \Eprint {http://arxiv.org/abs/astro-ph/0412641}
  {arXiv:astro-ph/0412641} \BibitemShut {NoStop}%
\bibitem [{\citenamefont {Abdo}\ \emph {et~al.}(2010)\citenamefont {Abdo} \emph
  {et~al.}}]{Abdo:2009ax}%
  \BibitemOpen
  \bibfield  {author} {\bibinfo {author} {\bibfnamefont {A.~A.}\ \bibnamefont
  {Abdo}} \emph {et~al.} (\bibinfo {collaboration} {Fermi-LAT}),\ }\href
  {\doibase 10.1088/0067-0049/187/2/460} {\bibfield  {journal} {\bibinfo
  {journal} {Astrophys. J. Suppl.}\ }\textbf {\bibinfo {volume} {187}},\
  \bibinfo {pages} {460} (\bibinfo {year} {2010})},\ \bibinfo {note} {[Erratum:
  Astrophys.J.Suppl. 193, 22 (2011)]},\ \Eprint
  {http://arxiv.org/abs/0910.1608} {arXiv:0910.1608 [astro-ph.HE]} \BibitemShut
  {NoStop}%
\bibitem [{\citenamefont {Gaensler}\ and\ \citenamefont
  {Slane}(2006)}]{Gaensler:2006ua}%
  \BibitemOpen
  \bibfield  {author} {\bibinfo {author} {\bibfnamefont {B.~M.}\ \bibnamefont
  {Gaensler}}\ and\ \bibinfo {author} {\bibfnamefont {P.~O.}\ \bibnamefont
  {Slane}},\ }\href {\doibase 10.1146/annurev.astro.44.051905.092528}
  {\bibfield  {journal} {\bibinfo  {journal} {Ann. Rev. Astron. Astrophys.}\
  }\textbf {\bibinfo {volume} {44}},\ \bibinfo {pages} {17} (\bibinfo {year}
  {2006})},\ \Eprint {http://arxiv.org/abs/astro-ph/0601081}
  {arXiv:astro-ph/0601081} \BibitemShut {NoStop}%
\bibitem [{\citenamefont {Zirakashvili}\ and\ \citenamefont
  {Aharonian}(2007)}]{Zirakashvili:2006pv}%
  \BibitemOpen
  \bibfield  {author} {\bibinfo {author} {\bibfnamefont {V.~N.}\ \bibnamefont
  {Zirakashvili}}\ and\ \bibinfo {author} {\bibfnamefont {F.}~\bibnamefont
  {Aharonian}},\ }\href {\doibase 10.1051/0004-6361:20066494} {\bibfield
  {journal} {\bibinfo  {journal} {Astron. Astrophys.}\ }\textbf {\bibinfo
  {volume} {465}},\ \bibinfo {pages} {695} (\bibinfo {year} {2007})},\ \Eprint
  {http://arxiv.org/abs/astro-ph/0612717} {arXiv:astro-ph/0612717} \BibitemShut
  {NoStop}%
\bibitem [{\citenamefont {{Reynolds}}\ \emph {et~al.}(2017)\citenamefont
  {{Reynolds}}, \citenamefont {{Pavlov}}, \citenamefont {{Kargaltsev}},
  \citenamefont {{Klingler}}, \citenamefont {{Renaud}},\ and\ \citenamefont
  {{Mereghetti}}}]{2017SSRv..207..175R}%
  \BibitemOpen
  \bibfield  {author} {\bibinfo {author} {\bibfnamefont {S.~P.}\ \bibnamefont
  {{Reynolds}}}, \bibinfo {author} {\bibfnamefont {G.~G.}\ \bibnamefont
  {{Pavlov}}}, \bibinfo {author} {\bibfnamefont {O.}~\bibnamefont
  {{Kargaltsev}}}, \bibinfo {author} {\bibfnamefont {N.}~\bibnamefont
  {{Klingler}}}, \bibinfo {author} {\bibfnamefont {M.}~\bibnamefont
  {{Renaud}}}, \ and\ \bibinfo {author} {\bibfnamefont {S.}~\bibnamefont
  {{Mereghetti}}},\ }\href {\doibase 10.1007/s11214-017-0356-6} {\bibfield
  {journal} {\bibinfo  {journal} {\ssr}\ }\textbf {\bibinfo {volume} {207}},\
  \bibinfo {pages} {175} (\bibinfo {year} {2017})},\ \Eprint
  {http://arxiv.org/abs/1705.08897} {arXiv:1705.08897 [astro-ph.HE]}
  \BibitemShut {NoStop}%
\bibitem [{\citenamefont {Fixsen}(2009)}]{Fixsen:2009ug}%
  \BibitemOpen
  \bibfield  {author} {\bibinfo {author} {\bibfnamefont {D.~J.}\ \bibnamefont
  {Fixsen}},\ }\href {\doibase 10.1088/0004-637X/707/2/916} {\bibfield
  {journal} {\bibinfo  {journal} {Astrophys. J.}\ }\textbf {\bibinfo {volume}
  {707}},\ \bibinfo {pages} {916} (\bibinfo {year} {2009})},\ \Eprint
  {http://arxiv.org/abs/0911.1955} {arXiv:0911.1955 [astro-ph.CO]} \BibitemShut
  {NoStop}%
\bibitem [{\citenamefont {Vernetto}\ and\ \citenamefont
  {Lipari}(2016)}]{Vernetto:2016alq}%
  \BibitemOpen
  \bibfield  {author} {\bibinfo {author} {\bibfnamefont {S.}~\bibnamefont
  {Vernetto}}\ and\ \bibinfo {author} {\bibfnamefont {P.}~\bibnamefont
  {Lipari}},\ }\href {\doibase 10.1103/PhysRevD.94.063009} {\bibfield
  {journal} {\bibinfo  {journal} {Phys. Rev. D}\ }\textbf {\bibinfo {volume}
  {94}},\ \bibinfo {pages} {063009} (\bibinfo {year} {2016})},\ \Eprint
  {http://arxiv.org/abs/1608.01587} {arXiv:1608.01587 [astro-ph.HE]}
  \BibitemShut {NoStop}%
\bibitem [{\citenamefont {Misiriotis}\ \emph {et~al.}(2006)\citenamefont
  {Misiriotis}, \citenamefont {Xilouris}, \citenamefont {Papamastorakis},
  \citenamefont {Boumis},\ and\ \citenamefont {Goudis}}]{Misiriotis:2006qq}%
  \BibitemOpen
  \bibfield  {author} {\bibinfo {author} {\bibfnamefont {A.}~\bibnamefont
  {Misiriotis}}, \bibinfo {author} {\bibfnamefont {E.~M.}\ \bibnamefont
  {Xilouris}}, \bibinfo {author} {\bibfnamefont {J.}~\bibnamefont
  {Papamastorakis}}, \bibinfo {author} {\bibfnamefont {P.}~\bibnamefont
  {Boumis}}, \ and\ \bibinfo {author} {\bibfnamefont {C.~D.}\ \bibnamefont
  {Goudis}},\ }\href {\doibase 10.1051/0004-6361:20054618} {\bibfield
  {journal} {\bibinfo  {journal} {Astron. Astrophys.}\ }\textbf {\bibinfo
  {volume} {459}},\ \bibinfo {pages} {113} (\bibinfo {year} {2006})},\ \Eprint
  {http://arxiv.org/abs/astro-ph/0607638} {arXiv:astro-ph/0607638} \BibitemShut
  {NoStop}%
\bibitem [{\citenamefont {Atwood}\ \emph {et~al.}(2009)\citenamefont {Atwood}
  \emph {et~al.}}]{Fermi-LAT:2009ihh}%
  \BibitemOpen
  \bibfield  {author} {\bibinfo {author} {\bibfnamefont {W.~B.}\ \bibnamefont
  {Atwood}} \emph {et~al.} (\bibinfo {collaboration} {Fermi-LAT}),\ }\href
  {\doibase 10.1088/0004-637X/697/2/1071} {\bibfield  {journal} {\bibinfo
  {journal} {Astrophys. J.}\ }\textbf {\bibinfo {volume} {697}},\ \bibinfo
  {pages} {1071} (\bibinfo {year} {2009})},\ \Eprint
  {http://arxiv.org/abs/0902.1089} {arXiv:0902.1089 [astro-ph.IM]} \BibitemShut
  {NoStop}%
\bibitem [{\citenamefont {Acero}\ \emph {et~al.}(2016)\citenamefont {Acero}
  \emph {et~al.}}]{Fermi-LAT:2016zaq}%
  \BibitemOpen
  \bibfield  {author} {\bibinfo {author} {\bibfnamefont {F.}~\bibnamefont
  {Acero}} \emph {et~al.} (\bibinfo {collaboration} {Fermi-LAT}),\ }\href
  {\doibase 10.3847/0067-0049/223/2/26} {\bibfield  {journal} {\bibinfo
  {journal} {Astrophys. J. Suppl.}\ }\textbf {\bibinfo {volume} {223}},\
  \bibinfo {pages} {26} (\bibinfo {year} {2016})},\ \Eprint
  {http://arxiv.org/abs/1602.07246} {arXiv:1602.07246 [astro-ph.HE]}
  \BibitemShut {NoStop}%
\bibitem [{\citenamefont {Ballet}\ \emph {et~al.}(2020)\citenamefont {Ballet},
  \citenamefont {Burnett}, \citenamefont {Digel},\ and\ \citenamefont
  {Lott}}]{Ballet:2020hze}%
  \BibitemOpen
  \bibfield  {author} {\bibinfo {author} {\bibfnamefont {J.}~\bibnamefont
  {Ballet}}, \bibinfo {author} {\bibfnamefont {T.~H.}\ \bibnamefont {Burnett}},
  \bibinfo {author} {\bibfnamefont {S.~W.}\ \bibnamefont {Digel}}, \ and\
  \bibinfo {author} {\bibfnamefont {B.}~\bibnamefont {Lott}} (\bibinfo
  {collaboration} {Fermi-LAT}),\ }\href@noop {} {\  (\bibinfo {year} {2020})},\
  \Eprint {http://arxiv.org/abs/2005.11208} {arXiv:2005.11208 [astro-ph.HE]}
  \BibitemShut {NoStop}%
\bibitem [{\citenamefont {Albert}\ \emph {et~al.}(2020)\citenamefont {Albert}
  \emph {et~al.}}]{Albert:2020fua}%
  \BibitemOpen
  \bibfield  {author} {\bibinfo {author} {\bibfnamefont {A.}~\bibnamefont
  {Albert}} \emph {et~al.} (\bibinfo {collaboration} {HAWC}),\ }\href {\doibase
  10.3847/1538-4357/abc2d8} {\bibfield  {journal} {\bibinfo  {journal}
  {Astrophys. J.}\ }\textbf {\bibinfo {volume} {905}},\ \bibinfo {pages} {76}
  (\bibinfo {year} {2020})},\ \Eprint {http://arxiv.org/abs/2007.08582}
  {arXiv:2007.08582 [astro-ph.HE]} \BibitemShut {NoStop}%
\bibitem [{\citenamefont {{Zimbardo}}\ \emph {et~al.}(1995)\citenamefont
  {{Zimbardo}}, \citenamefont {{Veltri}}, \citenamefont {{Basile}},\ and\
  \citenamefont {{Principato}}}]{1995PhPl....2.2653Z}%
  \BibitemOpen
  \bibfield  {author} {\bibinfo {author} {\bibfnamefont {G.}~\bibnamefont
  {{Zimbardo}}}, \bibinfo {author} {\bibfnamefont {P.}~\bibnamefont
  {{Veltri}}}, \bibinfo {author} {\bibfnamefont {G.}~\bibnamefont {{Basile}}},
  \ and\ \bibinfo {author} {\bibfnamefont {S.}~\bibnamefont {{Principato}}},\
  }\href {\doibase 10.1063/1.871453} {\bibfield  {journal} {\bibinfo  {journal}
  {Physics of Plasmas}\ }\textbf {\bibinfo {volume} {2}},\ \bibinfo {pages}
  {2653} (\bibinfo {year} {1995})}\BibitemShut {NoStop}%
\end{thebibliography}%

\clearpage

\end{document}